\documentclass{sig-alternate-05-2015}
\usepackage[utf8]{inputenc}
\usepackage{times}
\usepackage{helvet}
\usepackage{courier}
\usepackage{color}
\usepackage{xspace}
\usepackage{array} \usepackage{courier}

\definecolor{atk}{rgb}{0.698,0.122,0.435}
\definecolor{kwh}{rgb}{0.122,0.435,0.698}
\definecolor{cld}{rgb}{0.435,0.698,0.122}
\definecolor{todo}{rgb}{0.435,0.698,0.122}

\newcommand{\hashtag}[1]{{\small{\texttt{\#\MakeLowercase{#1}}}}}

\newcolumntype{L}{>{\centering\arraybackslash}m{5cm}}

\begin{document}
\title{Welcome or Not-Welcome: \\Reactions to Refugee Situation on Social Media}

\author{
  \alignauthor Asmelash Teka Hadgu, Kaweh Djafari Naini, Claudia Nieder{\'e}e\\
  	\affaddr{$ $}\\
  	\affaddr{L3S Research Center / Leibniz Universit\"{a}t Hannover, Germany}\\
  	\vspace{5pt}
    \email{\{teka, naini, niederee\}@L3S.de}
}

\maketitle

\begin{abstract}

For many European countries, in 2015 the refugee situation developed from a
remote tragedy reported upon in the news to a situation they have to deal with
in their own neighborhood. Driven by this observation, we investigated the
development of the perception of the refugee situation during 2015 in Twitter.
Starting from a dataset of 1.7 Million tweets covering refugee-related topics
from May to December 2015, we investigated how the discussion on refugees
changed over time, in different countries as well as in relationship with the
evolution of the actual situation. In this paper we report and discuss our
findings from checking a set of hypotheses, such as that the closeness to the
actual situation would influence the intensity and polarity of discussions and
that  news media takes a mediating role between  the actual and perceived
refugee situation.

\end{abstract}
 \section{Introduction}
\label{sec:intro}

The refugee situation in Europe has rapidly developed in the past year and is
further evolving making it a central topic in Europe.  The high numbers of
refugees in individual countries have created an unexpected readiness to help.
However, it also raises fears and unsureness, imposes demanding societal
challenges and triggers political developments. In each case, it has created a
high level of attention and discussion on the topic in the News as well as in
social media. While the perception of the refugee situation in the population
is reflected in the News, this typically rather highlights reactions in the
context of larger events than observing the evolution in a more continuous
fashion. 

Social media and especially Twitter have proven to be a useful source of
information in the context of events~\cite{Sakaki:2010:EST:1772690.1772777}. In
addition to the exchange of factual information they are also used to exchange
opinions on political and societal topics, especially in the context of larger
events.  Thus, they can become good indicators for the perception of and
opinion on events and evolving issues in the population. This has, for example,
been shown for the US presidential election in \cite{demartini11political}. The
tweeting activity on refugee-related topics, therefore, provides a good basis
for analysing the changing perception of the evolving refugee situation in
Europe. 

The reaction of the population to the arrival of large numbers of refugees in a
country is not easy to predict or analyse. Social science offers a variety of
theories for such intergroup encounter. The \textit{realistic group conflict
theory (RCT)} \cite{sherif1961intergroup,jackson93}, for example, has the
central hypothesis that real conflict in group interests between social groups
causes intergroup conflicts. In \cite{tajfel2004social} the authors stress the
importance of group identity in addition to RCT. Other driving factors in
intergoup relationships and conflicts are also prejudiced attitudes caused by
intraindividual and interpersonal psychological processes (see,
e.g.,\cite{berkowitz1962aggression}).  In addition to the variety of
influential factors for inter-group behavior, it has to be considered that the
population of a country is made up of a large variety of social groups
differing in their political attitude, life situation, education, employment
status, etc. Thus, for example, ``real conflicts in group interests'' -
contributing to intergroup conflicts according to RCT - such as competition for
employment will be perceived stronger by some social groups than others. 

In our work we analyse refugee-related tweets to gain a better understanding of
the perception of the evolving refugee situation in Europe.  For this purpose
we collected 1.7 Million tweets covering refugee-related topics from May to
December 2015. To the best of our knowledge this is the first work, which
analyses the current refugee situation and its perception based on Twitter
data.

In more detail, we test a set of hypotheses considering the influence of the
changing refugee situation on attention for the topic, critical to negative
attitude towards the topic, and polarization of the discussion. Since other media such
as online News play an important role as mediator between the actual refugee
situation and the one perceived by the population, we also
analyse the relationship between Twitter and online News activity based on the
News dataset provided by the GDELT project. This dataset has, for example, been used in
\cite{olteanu2015comparing} to assess the coverage of climate change on Twitter
and online News. 

The rest of the paper is structured as follows: Section~\ref{sec:relatedWork}
discusses related work followed by the presentation of our research hypotheses
in Section~\ref{sec:Hypothesis}. Subsequently, we describe the datasets in
Section \ref{sec:Dataset}, followed by a description of the methods applied for
processing the data to assess the analyses in Section \ref{sec:methods}. The
actual analysis results are presented in section~\ref{sec:analysis} and the
paper concludes with a summary of findings and ideas for future work
(Section~\ref{sec:Conclusions}). 
 \section{Related Work}
\label{sec:relatedWork}

\textbf{Sentiment Analysis on Twitter.} Due to its coverage and high level of
opinionated content Twitter is frequently used as source for analysing opinions
and feelings shared by people about different themes and topics.  Research has
leveraged the sentiments reflected in large-scale Twitter feeds for a variety
of purposes, such as predicting the stock market~\cite{Bollen20111}, political
alignments~\cite{Conover2010predicting} and
polls~\cite{Diakopoulos:2010:CDP:1753326.1753504}.  However, analysing
sentiments on Twitter using traditional methods for sentiment analysis such as
dictionaries is a challenging task~\cite{DBLP:conf/dis/BifetF10}, because
Twitter messages are typically very short, may contain positive and negative
sentiment in the same tweet, use a different language than other text corpora
and make use of special symbols such emoticons.
Some researchers have, therefore, exploited the special characteristics of
Twitter for sentiment analysis. 

In \cite{pak2010twitter} the authors created a dictionary for emoticons for
classifying tweets based on the employed emoticons. Further, they collected
objective tweets from Twitter accounts of newspapers such as the New York Times
and Washington Post to build a classifier for distinguishing tweets with
positive, negative and neutral sentiments. Recent works on sentiment analysis
also use the hashtags, another characteristic of Twitter, for identifying
sentiments and opinions. In \cite{Tsur:2012:WHC:2124295.2124320} hashtags are
employed to predict the spread of ideas in Twitter, while in
\cite{weber2013political} the authors use hashtags to explore polarizing US
political issues in Twitter. They compute the leaning of a hashtag by first
identifying users that retweet a set of ``seed users'' with a known political
leaning and then assigning each hashtag a fractional value corresponding to
which retweeting users used it. They remove non-political hashtags by requiring
hashtag co-occurrence patterns. Another study in
\cite{Wang:2011:TSA:2063576.2063726} investigates the task of sentiment
classification of hashtags instead of actual Twitter messages. For this
purpose, the authors leverage the overall sentiment polarity of tweets
containing a hashtag, hashtag co-occurrence relationship and the literal
meaning of the hashtags in a graph model framework. In our work, we also
exploit hashtags for sentiment classification basing the polarity of a hashtag
on its literal meaning and its usage. However, since we are not interested in
general polarity, but in polarity for a very specific topic, we decided to
resort to high-quality classification of the relevant hashtags using human
annotators. 

\textbf{Anlysing Events through Online Media.} 
Due to their accessibility and their ability for very dynamic 
reactions, online media including online news and social media 
are important sources for analysing events, their evolution, 
and their perception. 
In \cite{Utz201340}, for example, the authors analyse the effects of media type 
(Facebook vs. Twitter vs, online newspaper) and crisis type
(intentional vs. victim) on the reactions to information about events, crisis 
in this case, using the Fukushima Daiichi nuclear disaster as a 
use case. They show that the medium effects are stronger than the effects of
crisis type. In \cite{Schultz201120} the authors investigate theoretical foundations of
crisis communication by analyzing the effects of traditional and social media
strategies on the recipient's perceptions of reputation and by analyzing the
effects or crisis responses on the recipient's secondary crisis communications
(e.g., sharing information and leaving a message) and reactions (e.g.,
willingness to boycott). The results indicated that the medium matters more
than the message. In \cite{kwak2014first} the authors study global news coverage
of disasters using data from the GDELT project, which provides a dataset with
geolocated events of global coverage based on news reports from a variety of
international news sources \cite{Leetaru13gdelt:global}. In their study on news coverage
the authors examine predictor variables using a hierarchical multiple regression model.
They found that population has a positive effect, while political stability is
negatively correlated with global news coverage similar to previous research
but they find strong regionalism in news geography.
Similar to our work, the authors in \cite{olteanu2015comparing} 
also analyse a longer-lasting crisis situation, in this case 
climate change, based on Twitter data. The authors compare
the coverage of such events in online News and social media
and identified a gap between what the online
News covers and what the general public shares in social media.
Our focus is more on understanding the change in perception of the long-lasting 
situation over time and we focus on a different topic, namely the 
refugee situation. 

\textbf{Events and Crisis in Twitter.} The refugee situation in Europe has
developed a crisis like situation in many countries. Twitter has been
previously used to analyse crisis or event related aspects.
In~\cite{Sakaki:2010:EST:1772690.1772777} the authors use Twitter data to
collect and communicate information about earth quakes in Japan faster than the
announcements of the official meteorological agencies. Another challenge
during crisis is to identify false information such as rumours. A quantitative
analysis of tweets during the Ebola crisis reveals that lies, half-truths, and
rumours can spread just like true news~\cite{
DBLP:journals/computer/JinWZDCLR14}. Spiro et
al.~\cite{Spiro:2012:RDE:2380718.2380754} discussed different aspects of
classical rumour theory in the context of 2010 Deepwater Horizon oil spill by
analyzing the tweets during this event. The authors observed that an event with
high media coverage is more likely to be re-tweeted, meaning that perceived
importance is one major driver for the rumouring behaviour.
In~\cite{Olteanu:2015:EUH:2675133.2675242} the authors analyze a diverse set of
crisis situations such as floods, earth quake and train crashes in Twitter and
assess the usefulness of the data based on the information types and sources
for population, agencies and other stakeholders. Our work adds in the line of
crisis studies based on Twitter, introducing the refugee crisis as a new focus
topic.

 \section{Research Hypotheses}
\label{sec:Hypothesis}

The coverage of the refugee situation in the News suggests some trends such as
changes in the mood of the European population towards an increasing critical
attitude towards refugees, high attention of the population to the topic
(triggered by News coverage as well as by refugee presence in country), and
diverse perception of the topic in different countries (see e.g., UK and
Germany). Our goal is to verify some of those trends using Twitter data. Based
on this goal we have defined a set of research hypotheses, which we use for
guiding the analysis of the Twitter data on the refugee situation.
We have selected the following five hypotheses:
\begin{itemize}
    \setlength\itemsep{0em}
	\item Hypothesis I: The size and closeness of the refugee situation
        influences the intensity of perceiving and reflecting the topic;
	\item Hypothesis II: Media coverage acts as a mediator between the actual
        refugee situation and the reaction to it in the population;
	\item Hypothesis III: The critical attitude towards refugees increases with
        the number of refugees in the country;
    \item Hypothesis IV: The polarzation of discussion on refugee
        situation increases with the number of refugees in the country;
    \item Hypothesis V: There are considerable differences in the perception
        between different countries over time;
\end{itemize}

 \section{Dataset}
\label{sec:Dataset}

\textbf{Twitter Dataset on Refugees}. The basis of our analysis is a dataset of
tweets related to the topic refugees. Building upon the ideas presented in
\cite{ICWSM148091}, we used a multi-step approach for collecting our dataset.
We first collect a focused core dataset via a small set of keywords related to
the refugee situation. Subsequently, we use this dataset for harvesting a set
of relevant hashtags, which we use to collect additional tweets. In the third
step we combine the two datasets. Our process was designed in this way to keep
the crawl focused, avoiding the addition of non-relevant content. 

The focus of our collection process is on English and German content. With
English, we expect to get a good overview on the over-all refugee-related
content as 94\% of the crawled tweets are in English. The explicit addition of
German content is motivated by the fact that Germany is a key player as target
country in the refugee situation in Europe. We, therefore, expect that this
will give us a more complete picture. Furthermore, we use UK and Germany as
core countries in our analysis.

For collecting our dataset, we leveraged the Twitter advanced search
service\footnote{\url{https://twitter.com/search-advanced}} in combination with
the Twitter REST APIs. Our dataset generation begins with searching keywords
relevant for the topic: \textit{refugee(s)} for English and the equivalent
words \textit{fluechtling(e)} and \textit{fl\"uchtling(e}) for German.
Using the selected keywords, we performed queries on the Twitter advanced
search service to collect all public tweets containing any of these keywords
from May 1, 2015 to December 31, 2015. The resulting numbers of tweets and
unique contributing users are summarized in Table~\ref{tab:twitter_collection}.

From our keyword search result, some tweets, which are relevant for the topic,
might be missing just because they do not contain the selected keywords. To
improve recall, we performed another round of crawl using additional hashtags.
To keep our crawling process focused we decided to select only hashtags that
contain the word \emph{refugee} or the equivalent German word \emph{fluecht} or
\emph{fl\"ucht} for the second phase of the collection process. We considered
all hashtags with this pattern and comprise 99\% of the selected tweets. 
We used the advanced search interface to harvest all the tweets containing any
of the selected hashtags for the same time period as above. This helped us
mitigate the gap that is created by constraining tweets to contain the words
refugee(s) or their German translation. In addition, we didn't have to supply
the hashtags in advance. 

Finally, in the third step, we combined all unique tweets and users from both
the keyword search and hashtag search approaches. We used the Twitter REST API
to gather the complete json of each of these tweets, and the extended
information of the users. This way we obtain our final dataset (see
Table~\ref{tab:twitter_collection}).

\begin{table}
    \caption{Twitter collection counts.}
    \label{tab:twitter_collection}
    \centering
\begin{tabular}{l| r l l l}
    method   &  original  & retweets  & favorites & users    \\
    \hline
    keywords & 1,350,085  & 4,743,159 & 3,815,567 & 421,898  \\
    hashtags & 444,280    & 1,814,994 & 1,409,028 & 139,657  \\
    combined & 1,723,045  & 6,249,006 & 5,030,745 & 494,528  \\
\end{tabular}
\end{table}

\textbf{Refugees Count.} A \textit{refugee}, according to the Geneva Convention
on Refugees is a person who is outside their country of citizenship because
they have well-founded grounds for fear of persecution because of their race,
religion, nationality, membership of a particular social group or political
opinion, and is unable to obtain sanctuary from their home country or, owing to
such fear, is unwilling to avail themselves of the protection of that country.
Such a person may be called an \textit{asylum seeker} until considered with the
status of
\textit{refugee}\footnote{\url{https://en.wikipedia.org/wiki/Refugee}}.
We use the openly accessible data about asylum
applications\footnote{\url{http://popstats.unhcr.org/en/asylum_seekers_monthly}}
published by the UN Refugee Agency, UNHCR as an approximation to the number of
refugees. This dataset contains information about the number of asylum seekers
per month, the country where the asylum application was made, and the country
of origin of the asylum seekers. Such numbers can only be considered as an
approximation for the real number of refugees in Europe, since they are
typically incomplete and noisy. 
As \cite{crisp1999has} describes, the collection of accurate and consistent
refugee statistics is an extremely difficult task. A wide range of practical
obstacles hinder effective registration and counting. In addition, they might
also be influenced by the interests of host countries, countries of origin,
humanitarian agencies and other actors.
For example, the two million Syrian refugees in Turkey are not counted as
asylum seekers, but registered refugees. While the UNHCR publishes some data
about registered refugees, it is not as comprehensive as the data on asylum
seekers.
We therefore decided to use the number of asylum seekers as a reference for
assessing the size of the refugee situation in the individual countries in the
course of 2015.

\textbf{News Media Coverage.}
We use the dataset provided by the GDELT
project\footnote{\url{http://www.gdeltproject.org/}} to harvest information on
news coverage on the refugee topic. The GDELT Project provides a very large,
structured and annotated open spatio-temporal dataset covering online News. It
contains more than 302 million records in more than 300 categories dating back
to 1979 and is updated every 15 min. Since it uses several tens of thousands of
news sources from all over the globe, it can be considered as a good source for
assessing the news coverage on a topic at a given point in time. 
The GDELT project provides two types of raw datasets: (i) the GDELT Event
Database that contains records of events described in the world's news media
and (ii) the GDELT Knowledge Graph, which consists of two parallel data
streams, one encoding the entire knowledge graph with all of its fields and the
other that records ``counts'' of a set of predefined categories. We use the
later for our case. We looked at the GDELT category ``REFUGEE'' and take the
subset of the daily snapshots, which covers the period May to December 2015.
From this we have computed monthly numbers on News coverage on the topic
refugees. 
 \section{Methodology}
\label{sec:methods}

\subsection{Location} \label{sec:location}

Associating geo-location to tweets and users is an important task in order to
understand the spatio-temporal development of a topic, in our case the 
reflection of the refugee situation in Twitter in different countries.
There are two possible sources for location information, both imposing 
some challenges. First, tweets might be associated with 
geo-location reference. However, this information is scarce. Similar 
to results from other collections~\cite{Dredze_carmen:a},
only 1\% of the tweets in our collection have
geo-location associated with them. Second, Twitter users can put location
information into their profile. However, this location field is typically noisy. To overcome
these challenges, several approaches have been proposed to identify user
location information in Twitter \cite{Cheng:2010:YYT:1871437.1871535, Hecht:2011:TJB:1978942.1978976, TGIS:TGIS1297, Dredze_carmen:a}. 
The use of information from user profiles has been proven to provide the best results in terms of accuracy as presented in~\cite{Mahmud:2014:HLI:2648782.2528548}.

In our work we leverage the location information given by the user on their
profile for geo-locating tweets, since we are mostly interested in where the
user comes from and not so much in where the tweet was created. 
 We download the extended information of all 494,067 users in our collection
using the GET users/show REST
API~\footnote{\url{https://dev.twitter.com/rest/reference/get/users/show}}.
Users can put a short description in their profile with some basic information
including location information. In our collection 494,067 (76.68\%) users have
a non-blank location field. However, we can not use this field directly for
geo-location, because it is free text. This also means that not all content in
this field points to a valid location, e.g., location = 'wonderland'. Other
problems include people putting multiple valid locations, e.g., location =
'Berlin, san fransisco'. We use Yahoo! place maker to disambiguate the location
information. This way we found 301,542(61\%) of the users with a unique country
location. We use the ISO 3166-1 alpha-2 codes from
Wikipedia\footnote{\url{https://en.wikipedia.org/wiki/ISO_3166-1_alpha-2}} to
represent the countries in a standard unified way. These codes are also used in
place of the names of the countries throughout the paper.

\subsection{Polarized Tweets}\label{sec:opinion}
Applying traditional sentiment analysis methods on a collection of Tweets is
difficult due to the shortness and other characteristics of such microblogs.
Similar to \cite{weber2013political, Wang:2011:TSA:2063576.2063726}, we use
hashtags for identifying tweets that are opinionated with respect to the
refugee situation. The advantage in comparison with more traditional forms of
opinion mining is, that we can more precisely see the target of the opinion
(not only if a tweet is opinionated). 

We define the task of assigning a polarity to a hashtag to one of the three
categories: critical (-1), neutral (0) or positive (1) by looking into sample
tweets containing the hashtag in our collection and judging whether the
majority of tweets are, respectively, mostly negative, neutral or positive
towards the refugees, their coming to Europe and how this will influence the
future of the tweet poster's own country.

Since it is not feasible to annotate all hashtags, we selected top hashtags by
volume, using hashtags that appeared in at least 500 tweets. These were 302
hashtags that contribute to more than 80\% of the tweets in our collection. Not
to miss out polarized hashtags in the long tail, we also include 200 hashtags
that contain the word ``refugee'' and 100 hashtags that contained the
equivalent German word ``flucht''. These hashtags were selected by looking at
the top 99\% of hashtags containing their respective terms. All together, two
users annotated 554 unique hashtags into the three predefined categories. The
inter-annotator percentage agreement is 92\%. For the hashtags the annotators
disagreed, they jointly inspected the disagreements to generate a final
consensus. The annotation results form the basis on which we build the grouping
of tweets into negatively, positively and non-polarized tweets in
Section~\ref{sec:analysis}. Here are some examples of the annotated hashtags:

\noindent{\bf{Critical:}} \hashtag{refugeecrisis} \hashtag{norefugees}
\hashtag{nosyrianrefugees} \hashtag{refugeesnotwelcome}
\hashtag{norefugeeswelcome} \hashtag{nomorerefugees} \hashtag{nomuslimrefugees}
\hashtag{stoprefugees}

\noindent{\bf{Neutral:}} \hashtag{refugees} \hashtag{syrianrefugees}
\hashtag{refugee} \hashtag{syria} \hashtag{auspol} \hashtag{cdnpoli}
\hashtag{tcot} \hashtag{eunews} \hashtag{europe} \hashtag{migrants}
\hashtag{germany} \hashtag{greece} 
\noindent{\bf{Positive:}} \hashtag{refugeeswelcome} \hashtag{aidrefugees} \hashtag{helprefugees} \hashtag{refugeessafepassage}
\hashtag{refugeelifematters} \hashtag{refugeeconvoy} \\
\hashtag{solidaritywithrefugees} \hashtag{wewelcomerefugees}

\subsection{Weighting}

In most of our analysis, we experimented with various ways of counting
polarized hashtags by countries. For example we counted original tweets only,
incorporated retweets and favorites, or only counted unique contributing users. For most of
the analysis, the results are similar in trend and would yield the same conclusion. For
consistency, we stick with the count of original tweets, explicitly mentioning, 
where other numbers are used.

 \section{Analysis}\label{sec:analysis}

\subsection{Awareness and Attention}

For analysing Hypothesis I, we check how much attention is given to the refugee
topic depending upon the size and closeness of the refugee situation. We use
the number of European tweets per month on refugee topics as collected using
the methods described in Section~\ref{sec:Dataset} and the method detecting the
location described in Section~\ref{sec:location} as a proxy for the attention
given to the topic, i.e., the intensity with which the topic is perceived and
reflected by commenting on it. Furthermore, we use the number of asylum
applications of refugees coming to Europe in the respective months as issued by
the UN as a proxy for the size of the refugee situation in the proximity of
European citizens. 

\begin{figure}
    \centering
        \includegraphics[clip,, width=\linewidth]{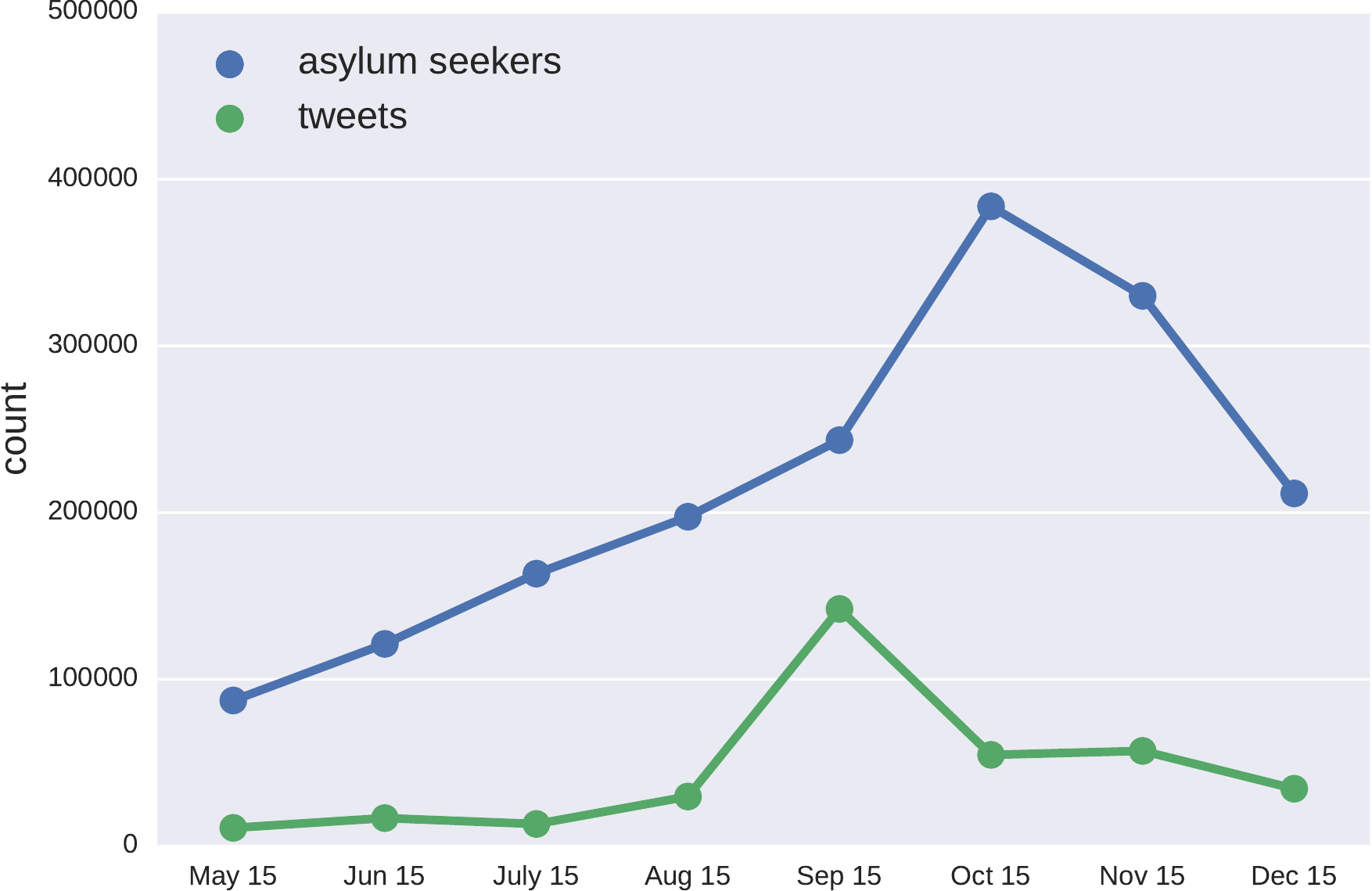}
    \caption{Twitter intensity (in terms of number of original tweets) and
    number of refugees (in terms of asylum applications of refugees coming to
Europe) May to December 2015.}
    \label{fig:Refugee_Number_and_Twitter_Intensity}
\end{figure}

Figure~\ref{fig:Refugee_Number_and_Twitter_Intensity} shows both the
development of the number of refugee-related tweets per month for Europe
together with the overall number of refugees coming to Europe on the same
timeline. When looking at the refugee numbers, we see a strong increase in the
numbers starting from less than 87,000 in May, a peak of more than 380,000 in
October and about 180,000 in December~\footnote{The drop of refugee numbers in
    December is partially due to the fact that not all the refugee numbers for
this month have been published yet.}. In addition, since more and more
refugees are arriving for considering the ''size'' of the refugee situation, it
also makes sense to look into the cumulative numbers, with strongly raising
numbers amounting to more than 1.7 Mio by the end of the year. 

If, in comparison, we take a look on the number of tweets, we see a peak in
autumn 2015. However, the peak is already in September 2015, thus preceding the
peak in the refugee numbers. In addition, the peak is followed by a fast
decrease in attention although the refugee numbers (and especially the
cumulative numbers) are still strongly increasing. This means that Hypothesis I
cannot be fully supported by the evidence we found in the data. Although,
overall the intensity of Twitter activity increases over the year, it shows a
different peaking behaviour. 

\begin{figure}[t!]
    \centering
        \includegraphics[clip,, width=\linewidth]{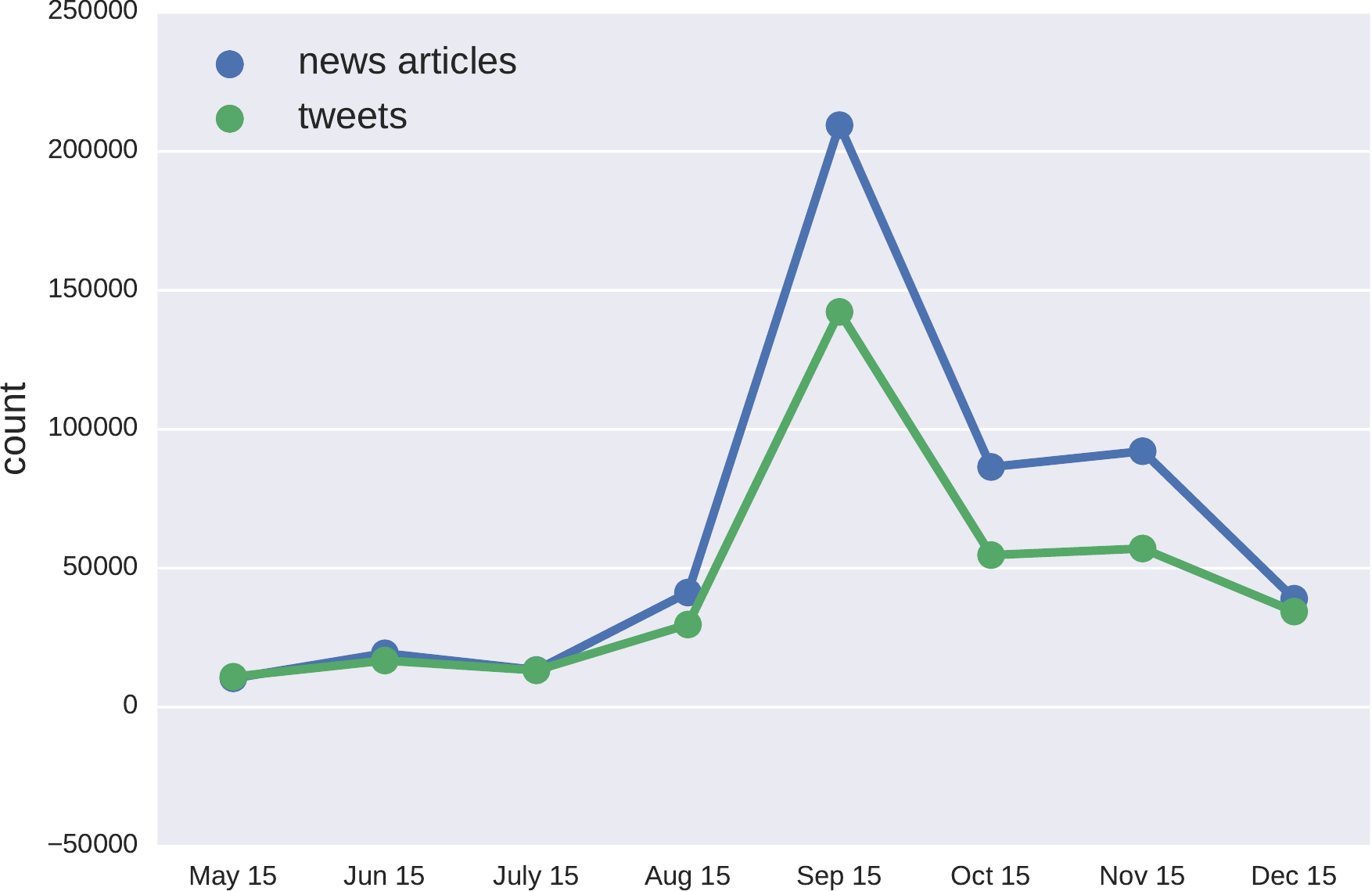}
    \caption{Twitter intensity in terms of tweets (number of original tweets)
    and News coverage (numbers of articles) May to December 2015.}
    \label{fig:News_and_Twitter_Intensity}
\end{figure}

This suggests that there is an additional factor that influences the tweet
activity on the refugee topic. News media is a candidate here, since it
strongly influences the perception of the refugee situation by the population
of a country. To check this secondary hypothesis, Hypothesis II, i.e., that
media coverage acts as a mediator between the actual refugee situation and the
reaction to it in Twitter, we compare the development of refugee-related
Twitter activity and News coverage from May to December 2015 (see
Figure~\ref{fig:News_and_Twitter_Intensity}). 

The numbers on News coverage are based on the classification and dataset
provided by the GDELT project (see also section \ref{sec:Dataset}). Looking at
those two timelines, their development (besides the actual altitude) is
strikingly similar. The development of news coverage and twitter intensity are
strongly correlated, which supports our Hypothesis II of news as a mediator
between the actual situation and its size, on the one hand, and the perceived
importance of the issue, which triggers twitter activity, on the other hand.

\subsection{Critical Attitude}

For checking Hypothesis III we first analyse the development of critical to
negative attitude towards refugees over time. We use the number of
refugee-related tweets classified as \textit{critical to negative} with the
methods described in section \ref{sec:opinion}.

\begin{figure}
    \centering
    \includegraphics[clip,, width=\linewidth]{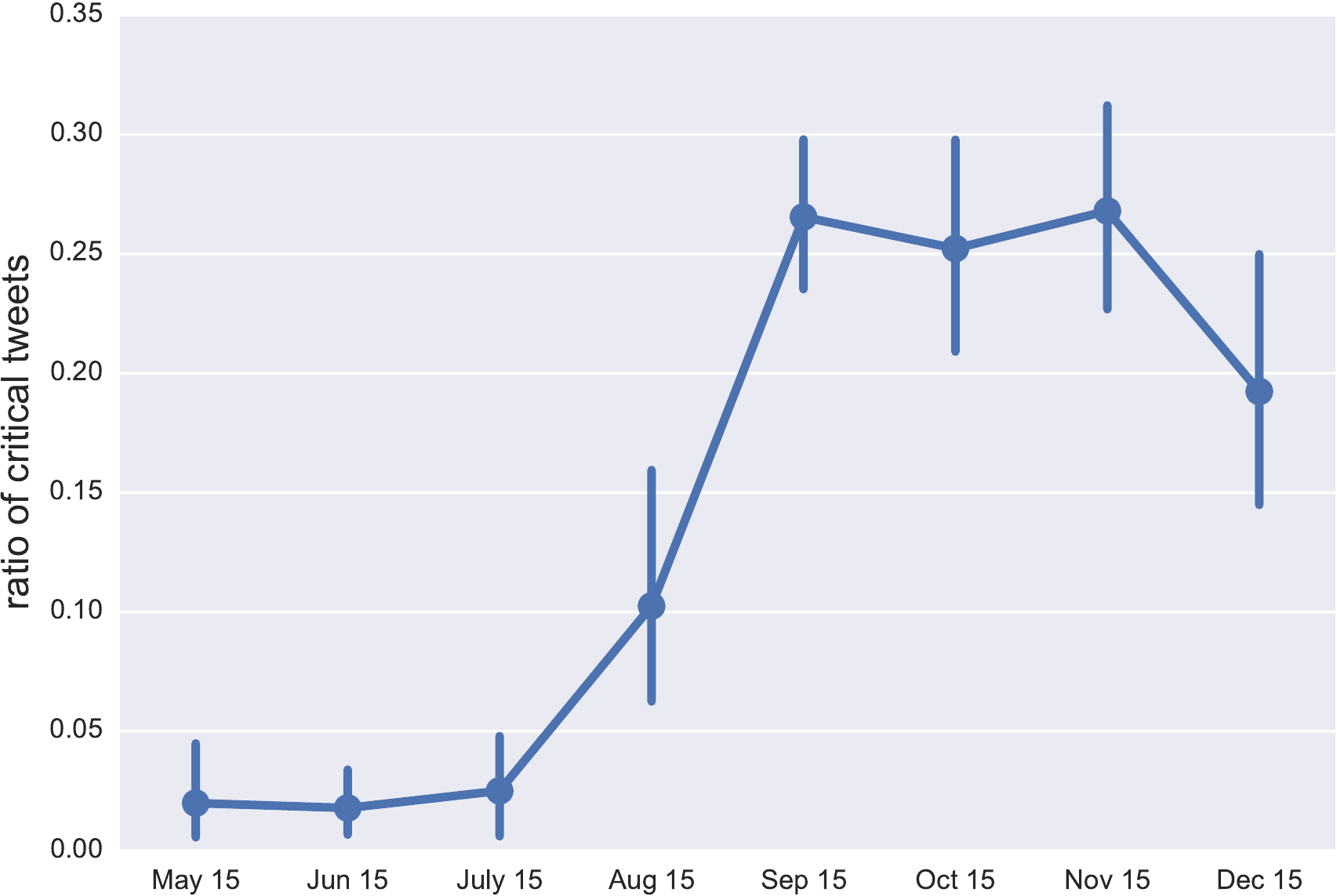}
    \caption{Percentage of Critical tweets to overall refugee-related tweets in
    Europe May to December 2015.}
    \label{fig:critical_negativeEU}
\end{figure}

Figure~\ref{fig:critical_negativeEU} shows the development of the percentage of
critical tweets over time for Europe. There is no constant increase of the
percentage of critical tweets over time, although the number of refugees grows
every month during this period of time. Rather, there is a strong increase in
the share of critical tweets in early autumn (August/September), when the
awareness for the refugee situation became very prominent, which reduces
towards the end of the year. Thus, Hypothesis III cannot be fully validated by
the data. However, the data shows that there is a strong increase in the rate
of the critical tweets as part of the overall tweets in the considered time
frame from less than 2\% in May and June to 20\% in December with peaks of more
than 25\% in September, October and November. 

\begin{figure}
    \centering
    \includegraphics[clip,, width=\linewidth]{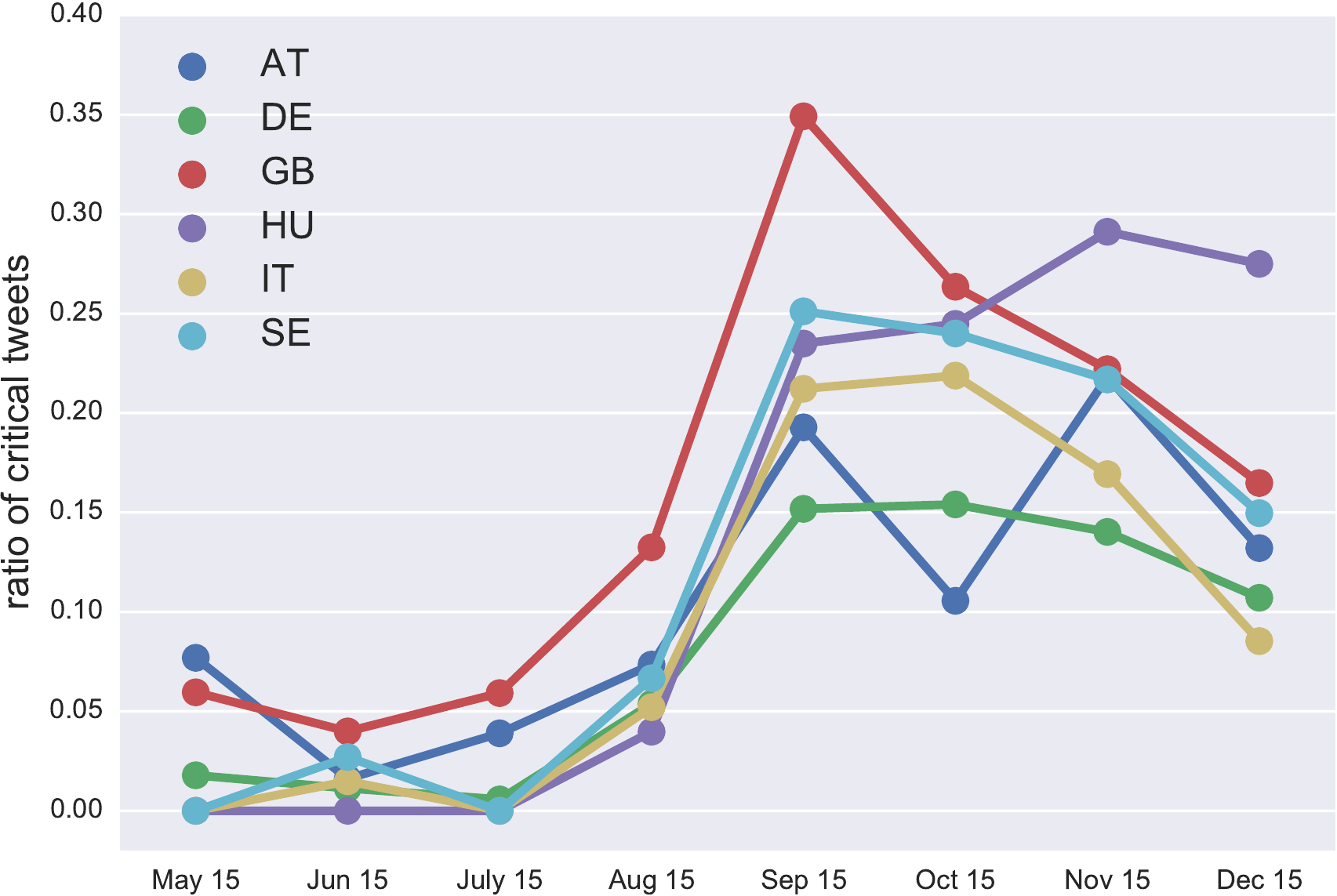}
    \caption{Percentage of Critical tweets to overall refugee-related
    tweets in selected Countries May to December 2015.}
    \label{fig:critical_negative}
\end{figure}

For a more detailed analysis, we looked into the data for selected European
countries (see Figure \ref{fig:critical_negative}). For this analysis, we
selected the five countries with the highest number of asylum applications in
the observation period. More precisely, we based our selection on the
accumulated numbers from May to November 2015, because the December numbers had
not yet been published by UNHCR for some of the European countries (such as UK,
Austria and Italy). For the selected countries, we still observe the general
trend. However, there is also considerable differences between the countries,
especially in the actual percentage of critical tweets e.g. between Germany and
Hungary.

Taking a closer look at actual refugee numbers in the individual countries, the
situation is even more diversified. For Hungary, for example, the refugee
numbers have dramatically dropped in October 2015 (from around 30 thousand in
September to less than 500 in October) due to closing of borders. However, the
share of critical tweets still increases until November with only a slight
relaxation in December. Furthermore, the share of critical tweets is also not
correlated to the actual number of newly arriving refugees in the countries.
UK, for example, with rather low numbers of asylum applicants is having very
high shares of critical tweets compared to other countries. In contrast,
Germany with very high refugee numbers has a relatively low share of negative
tweets. This results support hypothesis V that there are considerable
differences between the countries.

\begin{figure}[t!]
    \centering
    \includegraphics[clip,,width=\linewidth]{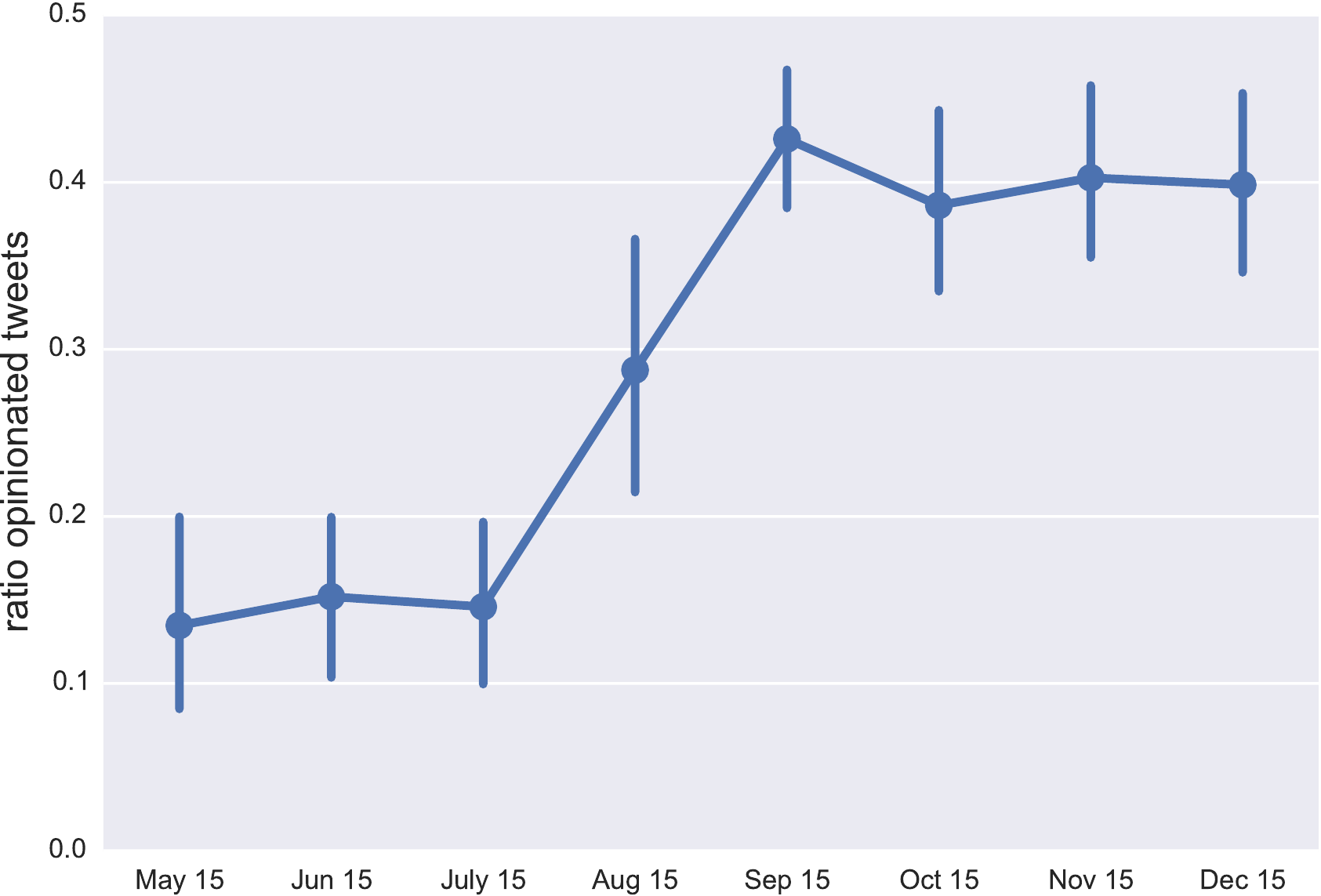}
    \caption{Percentage of Polarized tweets to overall refugee-related tweets in
    Europe May to December 2015.}
    \label{fig:critical_combinedEU}
\end{figure}

As a variant of Hypothesis III, we check Hypothesis IV, that the discussion on
the topic becomes more polarized with the increase in the size of the European
refugee situation. For analysing polarized tweets we aggregated the count of
positive and negative tweets (see Figure \ref{fig:critical_combinedEU}). We
observe an increase in the percentage during the year from less than 15\% to
about 40\% in December. We again observe a strong increase (a peak) in
September as in the case of the previous hypothesis. However the percentage
does not decrease afterwards. It stays on a high level until the end of the
year. 

\begin{figure}
    \centering
    \includegraphics[clip,, width=\linewidth]{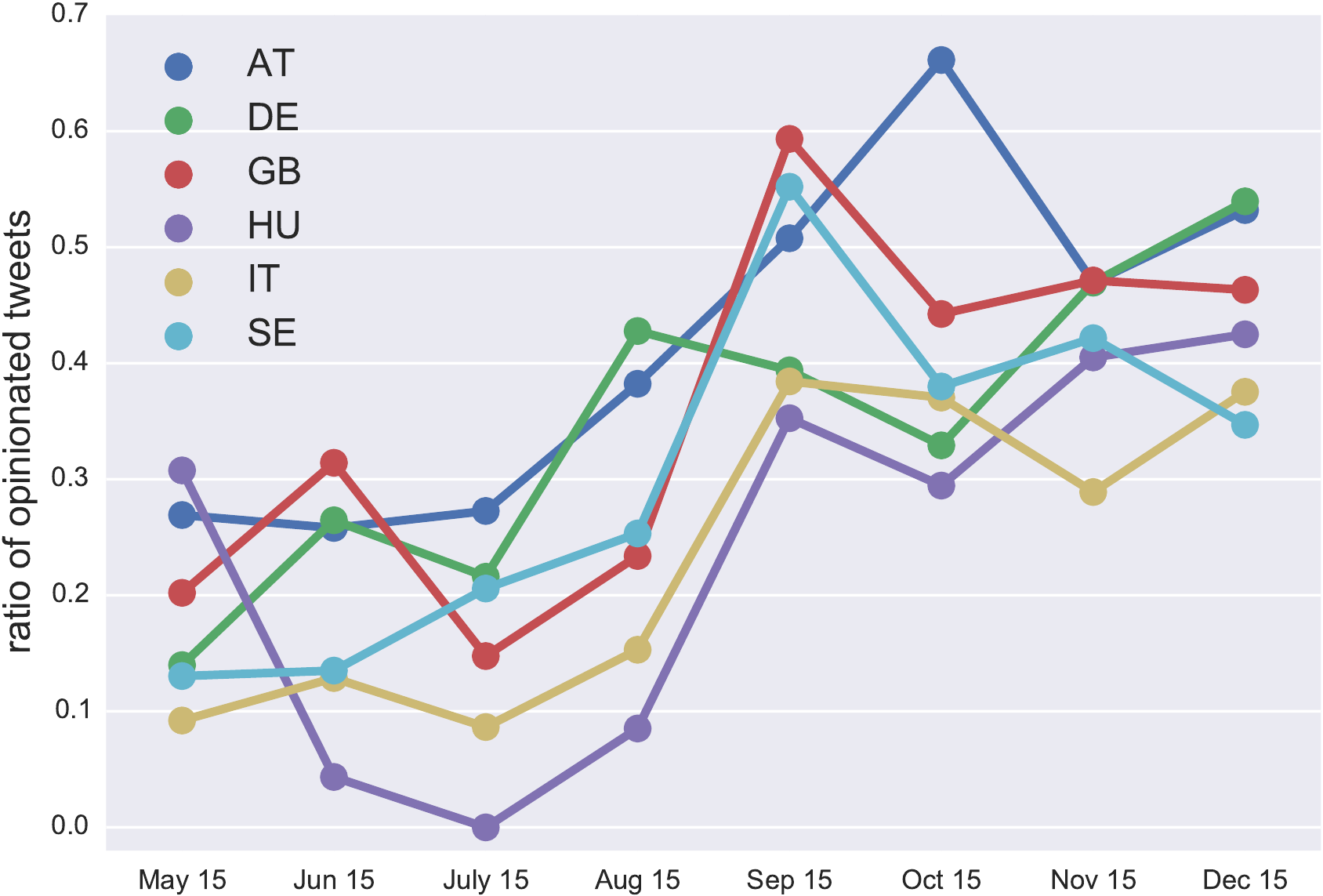}
    \caption{Percentage of Polarized tweets to overall refugee-related tweets
    for selected Countries May to December 2015.}
    \label{fig:critical_combined}
\end{figure}

If we look again on individual countries - using the same selected European
countries as before - (see Figure \ref{fig:critical_combined}), we see that for
most countries there is an upward trend in the percentage of polarized tweets
until the end of the year. However, there are some peaks probably due to
national politics and local events. The only exception are Serbia, which shows
a strong decrease in polarized tweets in December and UK, which shows a slight
decrease.  However, if we look at the refugee numbers for Serbia (in terms of
new asylum applications) they also show a strong decrease in Asylum
applications in the same time frame (from about 36k in November to 13k in
December 2015). Similarly, for UK there is a moderate decrease in the refugee
numbers from October to November (from nearly 5000 to about 3400).

Thus, to some degree the Hypothesis IV on an increase in polarized tweets with
increasing number of refugees in the respective country can be confirmed by our
data. With respect to hypothesis V also most countries follow a general upward
trend, however, there are considerable differences between them if you compare
the countries on the monthly bases (peaks). This confirms the hypothesis V for
this study.

 \section{Conclusion}
\label{sec:Conclusions}

In this paper we analysed the perception of the development of the refugee
situation in Europe from May to December 2015 as it is reflected in Twitter.
Starting from a refugee-related dataset, which we collected with a three step
approach, we analysed a set of five hypotheses. In our analysis, we could
confirm three of our hypotheses, namely the role of the News as a mediator
between the actual and the perceived refugee situation, the general trend for
an increase in polarized discussion with the increase in the number of refugees
in a country and the diversity of countries in the perception of the refugee
situation. The increase in Twitter activity and the increase of a critical to
negative attitude with a growing number of refugees, as it is suggested by
media, could not be fully confirmed.

The performed analysis raises a number of further questions, which could be
analysed in future work. One of them is a deeper analysis of the situation in
other countries besides Germany and UK, which would require to extend the
dataset to more languages. Also the inclusion of the refugees' voices into the
analysis would be very interesting. Furthermore, it would be interesting to
analyse the trends for subgroups of the population, e.g., by gender or other
demographic attributes getting deeper insights on the perception of the refugee
situation in such groups.

{
  \scriptsize
  \bibliographystyle{abbrv}
  \bibliography{bibliography}
}

\end{document}